\documentclass[aps,prx,amstext,amsmath,twocolumn,superscriptaddress]{revtex4-1}

\usepackage{graphicx}
\usepackage{amsmath}
\usepackage{color}
\usepackage{xspace}

\newcommand{\cak}{CaKFe$_4$As$_4$\xspace}

\begin{document}

\title{Analysis of the London penetration depth in Ni doped CaKFe$_4$As$_4$.}

\affiliation{Politecnico di Torino, Department of Applied Science and Technology, Torino 10129, Italy}
\affiliation{Istituto Nazionale di Fisica Nucleare, Sezione di Torino, Torino 10125, Italy}
\affiliation{Ames Laboratory, Ames, Iowa 50011, USA}
\affiliation{Department of Physics \& Astronomy, Iowa State University, Ames, Iowa 50011, USA}
\affiliation{National Research Nuclear University MEPhI (Moscow Engineering Physics Institute), Moscow 115409, Russia}

\author{D.~Torsello}
\affiliation{Politecnico di Torino, Department of Applied Science and Technology, Torino 10129, Italy}
\affiliation{Istituto Nazionale di Fisica Nucleare, Sezione di Torino, Torino 10125, Italy}
\affiliation{Ames Laboratory, Ames, Iowa 50011, USA}

\author{K.~Cho}
\affiliation{Ames Laboratory, Ames, Iowa 50011, USA}

\author{K.~R.~Joshi}
\affiliation{Ames Laboratory, Ames, Iowa 50011, USA}
\affiliation{Department of Physics \& Astronomy, Iowa State University, Ames, Iowa 50011, USA}

\author{S.~Ghimire}
\affiliation{Ames Laboratory, Ames, Iowa 50011, USA}
\affiliation{Department of Physics \& Astronomy, Iowa State University, Ames, Iowa 50011, USA}

\author{G.~A.~Ummarino}
\affiliation{Politecnico di Torino, Department of Applied Science and Technology, Torino 10129, Italy}
\affiliation{National Research Nuclear University MEPhI (Moscow Engineering Physics Institute), Moscow 115409, Russia}

\author{N.~M.~Nusran}
\affiliation{Ames Laboratory, Ames, Iowa 50011, USA}
\affiliation{Department of Physics \& Astronomy, Iowa State University, Ames, Iowa 50011, USA}

\author{M.~A.~Tanatar}
\affiliation{Ames Laboratory, Ames, Iowa 50011, USA}
\affiliation{Department of Physics \& Astronomy, Iowa State University, Ames, Iowa 50011, USA}

\author{W. R. Meier}
\affiliation{Ames Laboratory, Ames, Iowa 50011, USA}
\affiliation{Department of Physics \& Astronomy, Iowa State University, Ames, Iowa 50011, USA}

\author{M. Xu}
\affiliation{Ames Laboratory, Ames, Iowa 50011, USA}
\affiliation{Department of Physics \& Astronomy, Iowa State University, Ames, Iowa 50011, USA}

\author{S.~L.~Bud'ko}
\affiliation{Ames Laboratory, Ames, Iowa 50011, USA}
\affiliation{Department of Physics \& Astronomy, Iowa State University, Ames, Iowa 50011, USA}

\author{P.~C.~Canfield}
\affiliation{Ames Laboratory, Ames, Iowa 50011, USA}
\affiliation{Department of Physics \& Astronomy, Iowa State University, Ames, Iowa 50011, USA}

\author{G.~Ghigo}
\affiliation{Politecnico di Torino, Department of Applied Science and Technology, Torino 10129, Italy}
\affiliation{Istituto Nazionale di Fisica Nucleare, Sezione di Torino, Torino 10125, Italy}

\author{R.~Prozorov}
\affiliation{Ames Laboratory, Ames, Iowa 50011, USA}
\affiliation{Department of Physics \& Astronomy, Iowa State University, Ames, Iowa 50011, USA}

\date{\today}

\begin{abstract}

We report combined experimental and theoretical analysis of superconductivity in CaK(Fe$_{1-x}$Ni$_x$)$_4$As$_4$ (CaK1144) for $x=$0, 0.017 and 0.034. To obtain the superfluid density, $\rho=\left(1+\Delta \lambda_L(T)/\lambda_L(0) \right)^{-2}$, the temperature dependence of the London penetration depth, $\Delta \lambda_L (T)$, was measured by using tunnel-diode resonator (TDR) and the results agreed with the microwave coplanar resonator  (MWR) with the small differences accounted for by considering a three orders of magnitude higher frequency of MWR. The absolute value of $\lambda_L (T \ll T_c) \approx \lambda_L(0)$ was measured by using MWR, $\lambda_L (\mathrm{5~K}) \approx 170 \pm 20$ nm, which agreed well with the NV-centers in diamond optical magnetometry that gave $\lambda_L (\mathrm{5~K}) \approx 196 \pm 12$ nm.
The experimental results are analyzed within the Eliashberg theory, showing that the superconductivity of CaK1144 is well described by the nodeless s$_{\pm}$ order parameter and that upon Ni doping the interband interaction increases.
\end{abstract}

\maketitle

\section{Introduction}\label{S_intro}
The CaK(Fe$_{1-x}$Ni$_x$)$_4$As$_4$ (1144) family of iron-based superconductors (IBS) are particularly suitable for the studies of the fundamental superconducting properties due to the stoichiometric composition of the ``optimal" compound, CaKFe$_4$As$_4$, exhibiting clean-limit behavior and having a fairly high critical temperature, $T_c \approx 35 K$. This allows working with a system where unwanted effects caused by large amount of chemically substituted ions are minimal. Multiple experimental and theoretical results are compatible with the clean-limit nodeless s$_{\pm}$ symmetry of the order parameter in 1144 system and with all six electronic bands contributing to the superconductivity \cite{Cho2017, Teknowijoyo2018,Mou2016,Ummarino2016,Biswas2017,Fente2018}.\\
A rich and intriguing $T-x$ phase diagram emerges upon electron doping of the parent \cak, for example, by a partial substitution of Ni for Fe \cite{Budko2018}. The peculiarity of this system is that it exhibits an antiferromagnetic (AFM) state without nematic order (contrary to most IBS) called spin-vortex crystal (SVC) structure \cite{Meier2018,Ding2018}. Its presence is related to the existence of two nonequivalent As sites induced by the alternation of Ca and K as spacing planes \cite{Kreyssig2018} between the Fe-As layers that support superconductivity\cite{MazinPhysC}. It was suggested that a hidden AFM quantum-critical point (QCP) could exist in the CaK(Fe$_{1-x}$Ni$_x$)$_4$As$_4$ system near $x=0$ \cite{Ding2018}.\\ 
A very useful approach to investigate the pairing state of a material, the presence of nodes in its superconducting gaps and the presence of a QCP in its phase diagram, is to study the London penetration depth $\lambda_L$ and its changes in different compositions across the phase diagram \cite{Prozorov2011report}.
The low-temperature variation of the London penetration depth, $\Delta\lambda_L(T \ll T_c)$,  is directly linked to the amount of thermally excited quasiparticles. Exponential behavior is expected for a fully gapped Fermi surface and $T-$linear variation is obtained in the case of line nodes. Therefore, the analysis of the exponent, $n$, in the power-law fitting of the high-resolution measurements of $\Delta\lambda_L(T)=AT^n$ can be used to probe gap anisotropy, including the nodal gap \cite{,Prozorov2011report,SciAdvBaK122lambda2016,Cho2017,Cho2018SST}.
On the other hand, in the clean limit, the absolute value of the London penetration depth depends only on the normal state properties, notably the effective electron mass, 
$\lambda^2_L(0) \sim m^*$. Measurements of $\lambda_L(0)$ as a function of doping reveal a peak deep in the superconducting state due to the effective mass enhancement approaching a quantum phase transition \cite{Hashimoto_science,Almoalem2018,Wang2018,Joshi2019FeCo}.
Theoretically, London penetration depth can be computed on quite general grounds using the Eliashberg theory and, reproducing experimental data, one can discuss intrinsic quantities, such as the gap values and the coupling matrix coefficients \cite{Golubov,Torsello2019,Ghigo2017scirep,Ghigo2018PRL}.\\
In this work, a complete picture, from the experimentally determined superfluid density to Eliashberg analysis is obtained in CaK(Fe$_{1-x}$Ni$_x$)$_4$As$_4$ system for three different compositions, $x=$0, 0.017 and 0.034. To achieve this, we employed three complementary measurement techniques that combined provide a full and objective experimental information. Specifically, we used high$-T_c$-based microwave coplanar resonator (MWR), the tunnel diode resonator (TDR) and the NV-centers in diamond optical magnetometry. This systematic approach enabled us to discuss details of the pairing state and the most likely effect of Ni doping, in particular on the interaction matrix.\\
The paper is organized as follows. In Sect.\ref{S_exp} the experimental and theoretical techniques are explained, the results are presented and discussed in Sect.\ref{S_results} in terms of what can be deduced from them, and finally conclusions are drawn in Sect.\ref{S_conclusions}.

\section{Experimental techniques and theoretical methods}\label{S_exp}
\subsection{Crystals preparation}\label{S_prep}
High quality single crystals of CaK(Fe$_{1-x}$Ni$_x$)$_4$As$_4$ with doping levels of x=0, x=0.017 and x=0.034, were grown by high temperature solution growth out of FeAs flux. The Ni doping level was determined by wavelength-dispersive x-ray spectroscopy (for details of the synthesis and complete characterization see Ref. \cite{Meier2016}). All the investigated crystals were cleaved and reduced to the form of thin rectangular plates with thickness of less than 50 $\mu$m, in the direction of the $c$-axis of the crystals, and width and length one order of magnitude larger.

\subsection{Tunnel-diode resonator}\label{S_TDR}
\begin{figure}[h!]
\begin{center}
\includegraphics[keepaspectratio, width=\columnwidth]{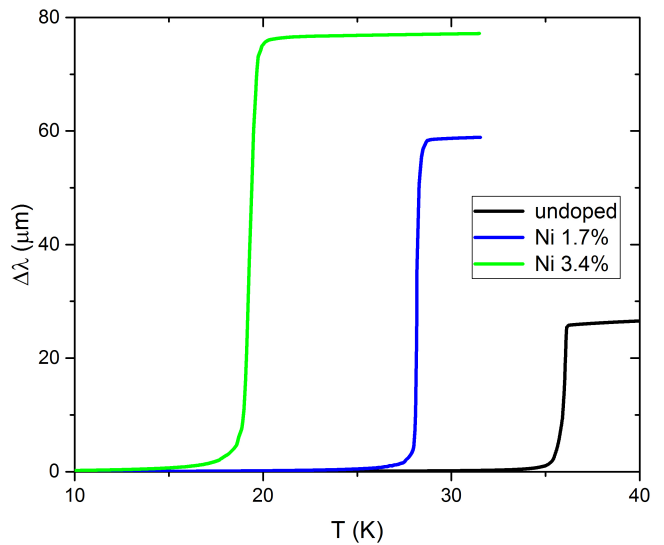}
\caption{Temperature dependence of $\Delta\lambda_{L,ab}$ for all doping levels measured with the TDR technique. The critical temperatures reported in Tab. \ref{tab:parameters} correspond to the maximum of the temperature derivative of these curves. The good quality and low disorder level of the samples is testified by the narrow transitions.}\label{Fig_transitions}
\end{center}
\end{figure}

A temperature variation of the London penetration depth in-plane component $\Delta\lambda_{L,ab}$(T) was measured using a self-oscillating tunnel-diode resonator (TDR) where the sample is subject to a small ac magnetic field parallel to the $c$-axis of the sample. This field configuration induces in plane supercurrents $j_{ab}$ and allows a direct measure of $\lambda_{L,ab}$. The resonant frequency shift from the value of the empty resonator is recorded and is  proportional to the sample magnetic susceptibility, determined by $\lambda_L$ and the sample shape, in the end the variations of the penetration depth with temperature $\Delta\lambda_{L,ab}(T)=\lambda_{L,ab}(T)-\lambda_{L,ab}(0)$ can be determined (see Fig. \ref{Fig_transitions}). A detailed description of this technique can be found elsewhere \cite{Prozorov2000PRB,Prozorov2006sust,Prozorov2011report,Cho2018SST}.

\subsection{NV centers magnetometry}\label{S_NV}
The determination of the low temperature absolute value of the London penetration depth $\lambda_L(0)$ is carried out by means of the newly developed NV centers magnetometry technique \cite{Joshi2019}. This consists in the measurement of  the field of the first vortex penetration $H_p$ on the sample edge by looking at the the optically detected magnetic
resonance (ODMR) of Zeeman-split energy levels in the NV centers of a diamond indicator positioned directly on top of the analyzed sample. From $H_p$ it is possible to calculate the value of the lower critical field $H_{c1}$ considering the effective demagnetization factor $N$ for a $2a$ $\times$ $2b$ $\times$ $2c$ cuboid in a magnetic field along the $c$ direction:
\begin{eqnarray}
H_p&&=H_{c1}(1+N\chi)\\
N^{-1}&&=1+\frac{3}{4}\frac{c}{a}\left(1+\frac{a}{b}\right)
\label{eq:NV}
\end{eqnarray}
where $\chi$ is the “intrinsic” magnetic susceptibility of the materialin the superconducting state, which can be taken to be equal to - 1.
Finally, from $H_{c1}$ it is possible to calculate the absolute value of the London penetration depth \cite{Hu1972}:
\begin{equation}
H_{c1}=\frac{\phi_0}{4\pi\lambda_L^2}\left(\ln\frac{\lambda_L}{\xi}+0.497\right).
\end{equation}
The coherence length $\xi$ can be calculated from the upper critical field and its uncertainty has a small influence on $\lambda_L(0)$ since it appears in the equation only as a logarithm.
\subsection{Microwave resonator}\label{S_MWR}
\begin{figure}[h!]
\begin{center}
\includegraphics[keepaspectratio, width=\columnwidth]{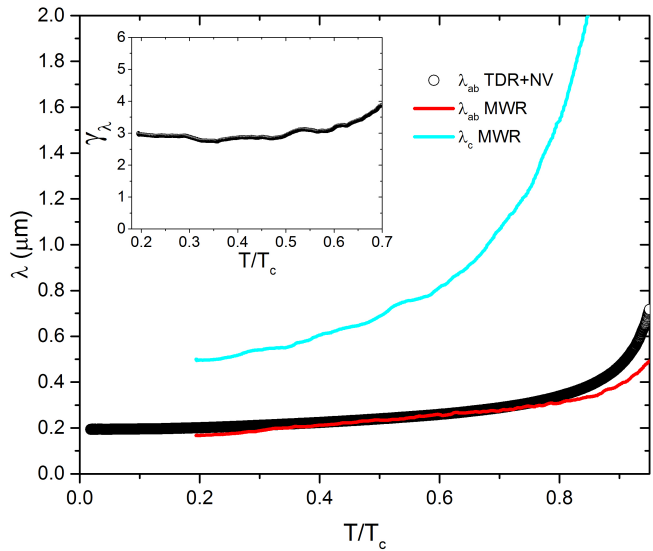}
\caption{Deconvolved components $\lambda_{L,ab}$ (red line) and $\lambda_{L,c}$ (cyan line) of the MWR measured $\lambda_L(T)$ for undoped CaKFe$_4$As$_4$. $\lambda_{L,ab}(T)$ from TDR+NV measurements is shown for comparison as black circles. The inset shows the anisotropy factor $\gamma_\lambda$.}\label{Fig_comparison}
\end{center}
\end{figure}
The complete characterization of the London penetration depth (absolute value and temperature dependence) can also be carried out by means of a microwave resonator (MWR) technique that has already been applied to other IBS crystals \cite{Ghigo2017prb,Ghigo2017scirep,Ghigo2018,Ghigo2018PRL,Torsello2019}. In this case the measurement system consists of an YBa$_2$Cu$_3$O$_{7-x}$ coplanar waveguide resonator (with resonance frequency $f_0$ of about 8 GHz) to which the sample is coupled. The whole resonance curve is recorded, making it possible to track not only frequency shifts but also variations of the quality factor, giving access to the absolute value of the penetration depth after a calibration procedure is performed.\\
It should however be noted that an important difference exists with respect to the TDR technique: in this case the applied ac magnetic field that probes the sample is oriented in plane instead of along the $c$-axis. For this reason the measurements yields an effective penetration depth $\lambda_L$ that is a combination of the main components $\lambda_{L,ab}$ and $\lambda_{L,c}$ dependent on the geometry of the sample under consideration.\\
In order to deconvolve the anisotropic contributions from the measured $\lambda_L$, one can study samples with different aspect ratios and analyze how they combine, considering that the penetration of the field occurs starting from all the sides of the crystal due to demagnetization effects from the sample that can not be considered infinite in any direction. The induced supercurrent is therefore in plane $j_{ab}$ in a thickness $\lambda_{L,ab}$ along the $c$-axis from both top and bottom faces, and out of plane $j_{c}$ in a thickness $\lambda_{L,c}$ along the $a$ and $b$ axes from the two sides. Accordingly, and in the hypothesis that $\lambda_{L,ab} \ll c$ and $\lambda_{L,c} \ll a, b$ (where c, a, b are respectively the thickness, width and length of the samples), the fraction of penetrated volume can be estimated as $\lambda_{L,ab}/c+\lambda_{L,c}/a+\lambda_{L,c}/b$ \cite{Ghigo2017prb}. Thus, the measured penetration depth can be expressed as:
\begin{equation}
\lambda_L=\lambda_{L,ab}+f_s \cdot \lambda_{L,c} ,
\label{eq.anisotropy}
\end{equation}
where $f_s=c\cdot(1/a+1/b)$ is the sample shape factor.\\
Considering two samples with different shape factors $f_s$, it is therefore possible to deconvolve the $\lambda_{L,ab}$ and $\lambda_{L,c}$ contributions to the total $\lambda_L$ measured, and to determine the anisotropy parameter $\gamma_\lambda=\lambda_{L,c}/\lambda_{L,ab}$. These quantities are shown in Fig. \ref{Fig_comparison} for the undoped samples. The substantial agreement between the $\lambda_{L,ab}$ curves validates the approach and the small differences will be discussed in Sect.\ref{S_comparison} in light of the differences between the TDR and MWR techniques. Lambda anisotropy, shown in the inset, is found to be comparable to that measured with $\mu$SR \cite{Khasanov2019}. Theoretically, substantial variation of the anisotropies of the characteristic lengths with temperature is consistent with multi-gap superconductivity and, as recently shown, can both increase or decrease with temperature depending on the order parameter symmetry and electronic structure \cite{Kogan2019}.
\subsection{Eliashberg modelling}\label{S_EE}
The experimental data can be reproduced within a two-bands Eliashberg s$_\pm$-wave model, allowing a deeper understanding of the fundamental properties of the material. The first step consists in calculating self consistently the gaps and the renormalization functions by solving the two-band Eliashberg equations, then from these quantities the London penetration depth can be calculated. The two-band Eliashberg equations \cite{Eliashberg,chubukov2008,bennemann2008superconductivity} are four coupled equations for the gaps $\Delta_{i}(i\omega_{n})$ and the renormalization functions $Z_{i}(i\omega_{n})$, where $i$ is a band index ranging from $1$ to $2$ and $\omega_{n}$ are the Matsubara frequencies. 
Starting from the general form of the Eliashberg equations, it is possible to reduce the number of input parameter by making some reasonable assumptions for the particular case under consideration. First of all one needs to identify the model for coupling that wants to consider. In the IBS it has been shown that electon-boson coupling is mainly provided by antiferromagnetic spin fluctuations \cite{MazinPhysC}, therefore we neglect the phononic contribution and we consider the shape of the spectral functions $\alpha^{2}_{ij}F^{sf}(\Omega)$ discussed in details in Refs. \cite{Torsello2019,Ghigo2017prb}. However, since the two-bands model is an effective one, it is not possible to set to zero the intraband coupling, hence the resulting electron-boson coupling-constant matrix $\Lambda_{ij}$ reads:
\begin{equation}
\vspace{2mm} %
\Lambda_{ij}= \left (
\begin{array}{ccc}
         \Lambda^{sf}_{11}                  &               \Lambda^{sf}_{12}            \\
              \Lambda^{sf}_{12}\nu_{12}               &                \Lambda^{sf}_{22}            \\
\end{array}
\right ) \label{eq:matrix}
\end{equation}
The parameter $\nu_{12}=N_{1}(0)/N_{2}(0)$ can be extracted from the ARPES measurements in Ref. \cite{Mou2016}, by assuming that the Fermi momentum in each band is proportional to the normal density of states at the Fermi level in the same
band, and adding the contribution of all hole bands for band 1 and all electron bands for band 2. The three constants $\Lambda^{sf}_{11}$, $\Lambda^{sf}_{22}$ and $\Lambda^{sf}_{12}$ will be free parameters (the only ones) of the model. \\
It is important to note that the choice of coupling mechanism limits the typology of order parameter obtainable, in the specific case of IBS and of AFM spin fluctuations the only order parameter symmetry allowed is $s_\pm$ \cite{Hirschfeld,MazinPhysC}. This specific state was chosen for the theoretical analysis because most experimental data points toward it \cite{Korshunov2018PRB,Hirschfeld2015PRB,Akbari2010PRB} and we find it is compatible with our data as well.\\
Once that the coupling mechanism has been defined, other terms of the general form of the Eliashberg equations can be set to zero: it has been shown that the Coulomb pseudopotential and the gap anisotropy can be neglected for IBS \cite{UmmarinoPRB2011,UmmarinoPRB2009,Hirschfeld}. Moreover, we decide to set to zero also the impurity scattering rate based on two observations: first the fact that the superconducting transition is very narrow for all samples indicates clean systems, second the effects of impurity scattering (increasing interband "mixing" and decreasing $T_c$) can be effectively considered by changing the coupling constants without adding free parameters in a simple two-bands effective model. It should be noted that this approach is only an effective one. The Ni atoms introduced in the structure are scattering centers, but their scattering potential can not be \textit{a priori} modelled within a simple scheme as in the case of irradiation induced disorder \cite{Ghigo2017scirep,Ghigo2018PRL}. For these reasons it is more convenient to practically take into account the effects of Ni doping by modifying the coupling matrix instead.\\
The imaginary-axis equations \cite{Eliashberg,Korshunov,UmmarinoPRB2011} under these approximations read: 
\begin{eqnarray}
&&\omega_{n}Z_{i}(i\omega_{n})=\omega_{n}+ \pi T\sum_{m,j}\Pi^{Z}_{ij}(i\omega_{n},i\omega_{m})N^{Z}_{j}(i\omega_{m})\phantom{aaa}
\label{eq:EE1}
\end{eqnarray}
\begin{eqnarray}
Z_{i}(i\omega_{n})\Delta_{i}(i\omega_{n})=\pi
T&&\sum_{m,j}\big[\Pi^{\Delta}_{ij}(i\omega_{n},i\omega_{m})\big]\nonumber\\
&&\times\Theta(\omega_{c}-|\omega_{m}|)N^{\Delta}_{j}(i\omega_{m}),\phantom{aaa}
 \label{eq:EE2}
\end{eqnarray}
with $\Pi^{Z}_{ij}(i\omega_{n},i\omega_{m})=\Lambda^{sf}_{ij}(i\omega_{n},i\omega_{m})$ and
$\Pi^{\Delta}_{ij}(i\omega_{n},i\omega_{m})=-\Lambda^{sf}_{ij}(i\omega_{n},i\omega_{m})$, where
\begin{equation}
\Lambda^{sf}_{ij}(i\omega_{n},i\omega_{m})=2\int_{0}^{+\infty}d\Omega \frac{\Omega\alpha^{2}_{ij}F^{sf}(\Omega)}{[(\omega_{n}-\omega_{m})^{2}+\Omega^{2}]}.
\end{equation}
\noindent $\Theta$ is the Heaviside function, $\omega_{c}$ is a cutoff energy and $sf$ stands for spin fluctuations. Moreover,
$N^{\Delta}_{j}(i\omega_{m})=\Delta_{j}(i\omega_{m})/
{\sqrt{\omega^{2}_{m}+\Delta^{2}_{j}(i\omega_{m})}}$ and
$N^{Z}_{j}(i\omega_{m})=\omega_{m}/{\sqrt{\omega^{2}_{m}+\Delta^{2}_{j}(i\omega_{m})}}$.
Finally, the electron-boson coupling constants are defined as
$\Lambda^{sf}_{ij}=2\int_{0}^{+\infty}d\Omega\frac{\alpha^{2}_{ij}F^{sf}(\Omega)}{\Omega}$.

The penetration depth can be computed starting from the gaps $\Delta_{i}(i\omega_{n})$ and the renormalization functions $Z_{i}(i\omega_{n})$ by
\begin{eqnarray}
&&\lambda^{-2}_L(T)=(\frac{\omega_{p}}{c})^{2}
\sum_{i=1}^{2}w^\lambda_i\pi T\notag\\
&&\times\sum_{n=-\infty}^{+\infty}\frac{\Delta_{i}^{2}(\omega_{n})Z_{i}^{2}(\omega_{n})}{[\omega^{2}_{n}Z_{i}^{2}(\omega_{n})+\Delta_{i}^{2}(\omega_{n})Z_{i}^{2}(\omega_{n})]^{3/2}}\label{eq.lambda}
\end{eqnarray}
where $w^\lambda_i=\left(\omega_{p,i}/\omega_{p}\right)^{2}$ are the weights of the single band contributions that sum up to 1 ($\omega_{p,i}$ is the plasma frequency of the $i$-th band and $\omega_{p}$ is the total plasma frequency). The multiplicative factor that involves the plasma frequencies derives from the fact that the low-temperature value of the penetration depth $\lambda_L(0)$ is related to the plasma frequency by $\omega_p=c/\lambda_L(0)$ for a clean uniform superconductor at $T=0$ if Fermi-liquid effects are negligible \cite{Korshunov}. In our specific case, we have only two additional free parameters: $w^\lambda_1$ and $\omega_{p}$.\\
The values of the remaining free parameters ($\Lambda_{11}$, $\Lambda_{22}$ and $\Lambda_{12}$) are set so that the experimental data is reproduced at best: gap values, critical temperature and temperature dependence of the superfluid density and London penetration depth. The procedure is the following.
The first step is to choose $\Lambda_{ij}$ values that, after solving self consistently Eqs. \ref{eq:EE1} and \ref{eq:EE2}, yield the critical temperature observed experimentally and low temperature values of the gap in agreement with those from tunneling measurements on similar undoped samples reported in \cite{Cho2017}. Then $\lambda_L(T)$ is calculated using Eq. \ref{eq.lambda}, and the superfluid density $\rho_s=(\lambda_L(0)/\lambda_L(T))^2$, that is independent of the $\omega_{p}$ value, is compared to the experimental one. During this step the value of the weight $w^\lambda_1$ is set to better compare with the experimental data. Finally, fine tuning of the $\Lambda_{ij}$ values is performed (the Eliashberg equations are solved again and penetration depth is recalculated until the best agreement with the experimental data is found). Then the $\omega_{p}$ value is set to obtain a $\lambda_L(0)$ value comparable to the experimental one.\\

\section{Results and discussion}\label{S_results}
\subsection{Techniques comparison}\label{S_tech}
Before focusing on the doping dependence of the penetration depth in the CaK(Fe$_{1-x}$Ni$_x$)$_4$As$_4$ system, we compare the results obtained for undoped CaKFe$_4$As$_4$ with the different experimental techniques described in Sect. \ref{S_exp}. The low temperature absolute values of the penetration depth $\lambda_L(0)$ from MWR and NV-centers magnetometry measurements of undoped CaKFe$_4$As$_4$ are remarkably close (170$\pm$20 nm and 196$\pm$12 nm respectively), considering that have been obtained with techniques that operate at different frequencies. Also the temperature dependence of $\lambda_{L,ab}$ from TDR and MWR shows an overall agreement (see Fig.\ref{Fig_comparison}) although different features emerge at low and high temperature. The deviation at high temperature is due to the fact that close to $T_c$ the deconvolution procedure of the MWR measurement looses its validity, because the assumption that  $\lambda_{L,ab} \ll c$ falls for the thinnest sample. The other deviation between the two measurement happens below $T/T_c$=0.3, where the $\lambda_{L,ab}$ MWR values dip lower, a feature that nicely corresponds to the increase of $\rho_s$ observed by Khasanov \textit{et al.} with the $\mu$SR technique \cite{Khasanov2019}. This difference could be explained by looking at the probing frequencies of the two techniques: we notice that at low temperature the characteristic time for pairbreaking scattering, that can be estimated within a two-fluids model as done in \cite{Ghigo2018}, becomes comparable to the characteristic time of the microwave probe ($\sim$ 125 ps) whereas the characteristic time of TDR is two orders of magnitude larger. It is therefore possible that the MWR techniques effectively eliminates the scattering contribution at low temperatures, resulting in a cleaner system with lower $\lambda_L$ values. The same argument applies to the comparison between the $\lambda_L(0)$ values measured with the MWR and NV-centers magnetometry techniques.

\subsection{Low temperature data}\label{S_LowT}
As stated in the introduction, from the low temperature behavior of  $\lambda_L$ it is possible to get important information about the pairing state of a superconducting material and, by carrying out a study along the phase diagram, also about the possible presence of a QCP.\\

\begin{figure}[h!]
\begin{center}
\includegraphics[keepaspectratio, width=\columnwidth]{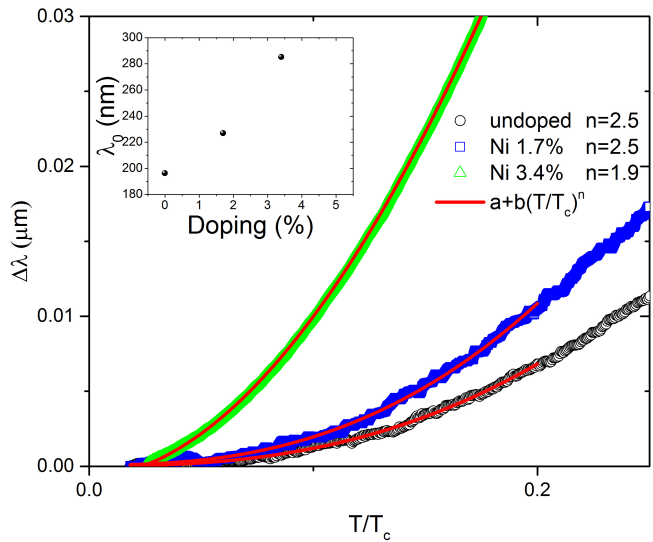}
\caption{Low temperature variation of the London penetration depth in CaK(Fe$_{1-x}$Ni$_x$)$_4$As$_4$ for all doping levels and their power-law fit $\Delta\lambda_L = a + b(T/T_c)^n$. $n = 2$ represents the dirty-limit exponent for the sign-changing order parameters $s_{\pm}$. The inset shows the $\lambda_L(0)$ values as a function of Ni doping.}\label{Fig_exponent}
\end{center}
\end{figure}

For each sample we fit the $\Delta\lambda_L(T)$curve with the exponential function $a+b(T/T_c)^n$ (see Fig. \ref{Fig_exponent}) up to a reduced temperature $t=T/T_c$=0.2 and discuss the possible presence of line nodes in the superconducting gaps in light of the obtained $n$ values. $n$=1 implies the gap has d-wave-like line nodes, exponential low-temperature behavior of $\lambda(T)$, mimicked by a large exponent $n$ $>$ 3-4, is expected for clean isotropic fully-gapped superconductors, and $n$ approaches 2 in dirty line-nodal (e.g., d-wave) and dirty sign-changing s$_\pm$ superconductors. \cite{Hirschfeld1993}. We find that $n$ decreases from 2.5 for the undoped sample to 1.9 for the $x$=0.034 sample, a strong indication that the system is fully gapped and that Ni doping increases disorder driving the system from the clean to the dirty limit. Moreover, it should be noted that a conventional BCS exponential behavior of $\Delta\lambda_L(T)$ is expected if the exponent $n \geq 3$. This is not the case for this system mainly due to the fact that scattering in s$_\pm$ superconductors is pair-breaking and that it presents multi-gap superconductivity. For these reasons the behavior can look conventional only below a temperature determined by the smallest gap, and much smaller than the usual $T_c$/3 threshold. It follows that the data shown in Fig. \ref{Fig_exponent} can not be fit well by a simple  generalized two s-wave gap scheme where $\Delta\lambda_L \propto \sum_i \lambda_L(0)\sqrt{(\pi/2|\Delta_i(0)|/(k_BT_ct))}\times\exp(-|\Delta_i(0)|/(k_BT_ct))$ \cite{Hashimoto2009PRL2}.\\
It is in principle possible to identify a QCP in the phase diagram by analyzing the $\lambda_L(0)$ curve as a function of doping level $x$: it would correspond to a sharp peak in the $\lambda_L(0)(x)$ plot \cite{Hashimoto_science,Almoalem2018}. In the present case such a feature is not visible (as evident from the inset in Fig. \ref{Fig_exponent}) due to the fact that a finer spacing in $x$ would be needed and/or to an effect of disorder that induces an increase of $\lambda_L(0)$ that in turn hides the QCP peak. This does not necessarily exclude the presence of QCP in the analyzed doping range.\\

\subsection{Eliashberg analysis}\label{S_comparison}
\begin{figure}[h!]
\begin{center}
\includegraphics[keepaspectratio, width=\columnwidth]{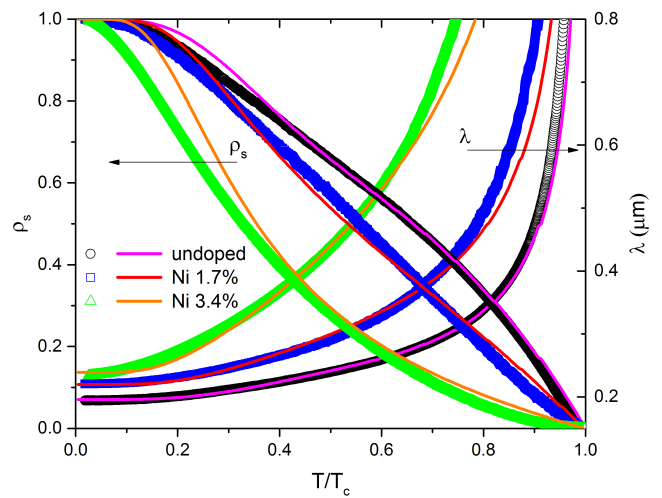}
\caption{Experimental $\rho_s$ and $\Delta\lambda_L$ vs T curves for all doping levels (black circles for the undoped sample, blue squares for 1.7\% Ni and green triangles for 3.4\%) compared to the results of the Eliasberg calculations shown as solid lines.}\label{Fig_fit}
\end{center}
\end{figure}

The Eliashberg equations were solved and the London penetration depth was calculated for all doping values following the approach explained in Sect.\ref{S_EE}, yielding the $\rho_s$ and $\Delta\lambda_L$ vs $T$ curves presented in Fig. \ref{Fig_fit} where they are compared to the experimental ones.\\

\begin{figure}[h!]
\begin{center}
\includegraphics[keepaspectratio, width=\columnwidth]{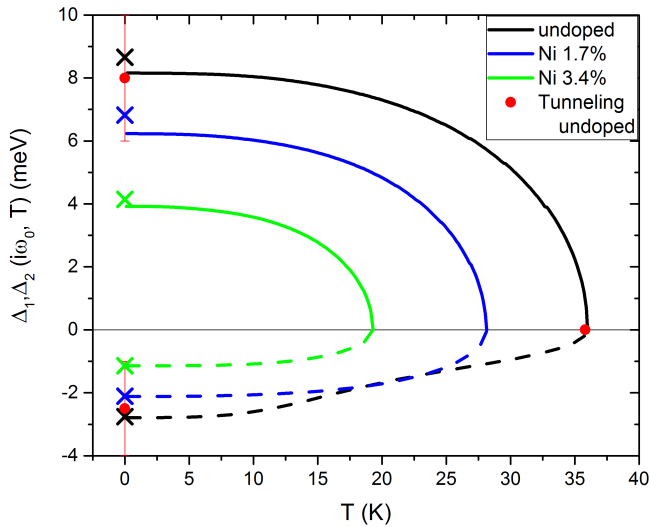}
\caption{Temperature dependence of the first value of the energy gaps for the investigated compounds obtained by the solution of the imaginary-axis Eliashberg equations. Crosses give the values obtained by analytical continuation on the real axis with the Pad\'e approximants. Experimental data for the stoichiometric composition (from \cite{Cho2017}) are shown as red circles for comparison.}\label{Fig_gaps}
\end{center}
\end{figure}

The excellent overall agreement, in particular considering that the model employed is an effective two-bands one, testifies that the s$_{\pm}$ symmetry is consistent with the observed data. The parameters used in the calculation are given in Tab. \ref{tab:parameters} and Fig. \ref{Fig_gaps} shows the calculated temperature dependence of the gaps for all doping values. The gap values obtained by analytical continuation on the real axis at low temperature (crosses in Fig. \ref{Fig_gaps}) for the undoped case are in nice agreement with those measured by the tunneling conductance technique in similar samples \cite{Cho2017}. With increasing doping (and therefore decreasing $T_c$) the gaps become smaller as expected. It is worth noticing that the shape of the small gap changes drastically between the undoped and doped samples, becoming more BCS-like when Ni substitutes Fe. This is due to an increase of the interband coupling ($\Lambda_{12}$ in Tab. \ref{tab:parameters}) necessary to reproduce the experimental $\rho_s$. This means that it is the large gap that determines the overall behavior of the system when Ni is introduced: chemical substitution increases scattering that intermixes more the bands, an effect that can be taken into account by either increasing interband scattering or effectively by changing the interband coupling.

\begin{widetext}
\begin{center}
\begin{table}[h!]
\caption{Summary of the experimental values and of the main model parameters used to reproduce the experimental data. $T_c$ is the experimental critical temperature, $\lambda_L(0)$ is the low-temperature penetration depth determined by NV magnetometry measurement, $\Lambda_{ij}$ are the coupling-constants for the Eliashberg equations, $\Delta_i$ are the low-temperature values of the gaps on the real axis, $\omega_p$ is the plasma frequency.  \label{tab:parameters}}
\setlength{\tabcolsep}{17pt}
\begin{tabular}{c|c c c c c c c c }
\hline 
\hline 
Ni doping & $T_c$ & $\lambda_L(0)$ &  $\Lambda_{11}$ &  $\Lambda_{22}$ &  $\Lambda_{12}$ & $\Delta_1$ & $\Delta_2$ & $\hbar \omega_p$  \\ 

\% & K & nm & • & • & • & meV & meV & meV  \\ 
\hline 
\hline 
0 & 36.0 & 196.4 & 0.80 & 2.77 & -0.10 & -2.76 & 8.66 & 10.3  \\ 
\hline 
1.7 & 28.2 & 227.0 & 0.10 & 2.34 & -0.30 & -2.11 & 6.82 & 7.7  \\ 
\hline 
3.4 & 19.3 & 285.2 & 0.00 & 1.51 & -0.22 & -1.14 & 4.14 & 5.8  \\ 
\hline 
\hline 

\end{tabular} 
\end{table}
\end{center}
\end{widetext}

\section{Conclusions}\label{S_conclusions}
In summary, we employed a combination of three experimental techniques (TDR, NV magnetometry and MWR) together with theoretical modelling based on the solution of the two-bands Eliashberg equations to demonstrate that a complete characterization of the London penetration depth allows to study in depth the fundamental properties of  superconducting materials. 
The comparison between the techniques on undoped CaKFe$_4$As$_4$ shows very good agreement both regarding the $\lambda_L(0)$ absolute values (170$\pm$20 nm for MWR and 196$\pm$12 nm for NV) and the $\lambda_{L,ab}$ temperature dependence. We ascribe small differences at low temperature to the high probe frequency of the microwave resonator technique that hinders pairbreaking scattering.
Overall, the CaK(Fe$_{1-x}$Ni$_x$)$_4$As$_4$ system (with doping levels between x=0 and x=0.034) shows properties compatible with the s$_{\pm}$ order parameter symmetry without line nodes in the gaps. Upon doping the system presents a stronger interdependence between the two bands, probably caused by scattering induced by the Fe-Ni substitution. No sign of a QCP was found in the variation of the penetration depth low temperature value upon doping, due to the low number of available data and possibly to the fact that the disorder induced increase of $\lambda_L(0)$ hides the QCP peak.

\begin{acknowledgments}
Work in Ames was supported by the U.S. Department of Energy, Office of Basic Energy Science, Division of Materials Sciences and Engineering. Ames Laboratory is operated for the U.S. Department of Energy by Iowa State University under Contract No. DE-AC02-07CH11358. D.T. thanks R.P., Ames Laboratory and Iowa State University for the opportunity of participating in the measurements reported in this paper at their facilities. G.A.U. acknowledges the support from the MEPhI Academic Excellence Project (Contract No. 02.a03.21.0005).
\end{acknowledgments}
\end{document}